\documentclass[preprint]{aastex}

\usepackage{emulateapj5}
\usepackage{epsfig}

\begin{document}

\title{The impact of a close stellar encounter on the Edgeworth-Kuiper Belt}

\author{Alice C. Quillen} 
\affil{Department of Physics and Astronomy,
University of Rochester, Rochester, NY 14627}
\email{aquillen@pas.rochester.edu}
\author{David E. Trilling, \&} 
\affil{University of Pennsylvania, 209 South 33rd Street, Philadelphia, PA 19104}
\email{trilling@hep.upenn.edu}
\author{Eric G. Blackman}
\affil{Department of Physics and Astronomy,
University of Rochester, Rochester, NY 14627}
\email{blackman@pas.rochester.edu}

\begin{abstract}
We numerically investigate the possibility 
that a close stellar encounter could
account for the high inclinations of the Kuiper belt, as originally proposed
by Ida, Larwood and Burkert, 
however we consider encounters with pericenters 
within those explored by previous works.
A star of mass $0.2 M_\odot$ entering the solar system
on a nearly parabolic, low inclination, retrograde 
orbit with perihelion of 50~AU can scatter $\sim 30\%$ of the Kuiper belt
into moderate inclination ($i>10^\circ$)  
and eccentricity orbits, while leaving
the rest of the belt at low eccentricities and inclinations.
This scenario provides a possible explanation for the dichotomy of
the Kuiper belt, accounting for the scattered and classical populations,
including objects with high eccentricities and inclinations.  
High eccentricity objects that were
not placed into resonance with Neptune are more likely to be
removed from the system by subsequent long timescale evolution. 
Consequently this scenario suggests that some Plutinos could have 
originally been objects in circular orbits 
which were scattered into resonant high eccentricity and high
inclination orbits by the stellar encounter.
Objects could have been placed into stable resonant regions with Neptune by
the encounter.  However long lived regions of dynamical stability
should differ from those selected by resonance capture.
This scenario predicts the presence of objects in resonances
with Neptune that are unlikely to have been captured via Neptune's migration.

\end{abstract}


\section{Introduction}

The Edgeworth Kuiper Belt (KB) is often discussed in terms of three
dynamical families; 1) the classical Kuiper Belt objects 
which occupy semi-major axes $41 \lesssim a \lesssim 46$ AU and which
have low eccentricities $e< 0.25$, 2) the resonant Kuiper Belt  
occupying the 3:2 and 2:1 mean motion resonances with Neptune, 
denoted the Plutinos and Twotinos, respectively, and 
3) the scattered Kuiper belt
occupying extreme orbits with high eccentricities ($e$ up to 0.6) 
(e.g, \citealt{trujillo2002}).
Most of the mass in the Kuiper belt appears to reside 
in the classical Kuiper belt \citep{luu}.  All three dynamical
families contain objects with surprisingly 
high inclinations ($i > 15^\circ$; \citealt{trujillo2002,luu,brown}).
Current compilations suggest a bimodal distribution, with $\sim 20\%$ 
of the Kuiper belt objects in low-inclination orbits 
of a few degrees and the remainder in highly inclined orbits with 
$i \gtrsim 20^\circ$ \citep{brown}.
Scenarios for the formation and evolution of the Kuiper Belt should explain 
the different Kuiper Belt Object (KBO) populations.

Several mechanisms have been proposed to account for the high 
KBO inclinations, including scattering by Neptune during
Neptune's migration \citep{gomes,levison}, passage through
vertical resonances during Neptune's migration or during the depletion
of the solar nebula \citep{nagasawa2000,malhotra98}, 
wave excitation \citep{ward},
perturbations from passing stars \citep{ida,kobayashi}, and
scattering by planets and planetesimals \citep{thommes,petit,kenyon,brunini}.
While long term evolution can remove KBOs with low inclinations,
it cannot explain the existence of classical KBOs 
with extremely high inclinations
of $30^\circ$ \citep{kuchner}.  Current surveys appear to rule out
the possibility of a distant planetary embryo \citep{morbidelli}.

In this paper we focus on the possibility that perturbations
from a stellar encounter could account for the high KBO inclinations, 
as proposed by \citet{ida}.
The probability of a stellar encounter that passes with a few
hundred AU of the Sun at its current position is extremely unlikely. 
However, most stars are formed in stellar groups or clusters \citep{lada}, 
associated for $10^7 - 10^9$ years until they are tidally disrupted by molecular
clouds and structure 
in the Milky Way disk \citep{lada,terlevich,theuns,bergond}.  
Young stellar groups and clusters, can have stellar densities 
up to $10^5$ times higher than the Solar neighborhood \citep{lada,hillenbrand}. 
The Sun could have experienced a close stellar encounter
when it was still in its birth cluster, 
\citep{ida,adams,gaidos}.
Other disk systems such as Beta Pictorus and HD 100546 probably have
suffered close stellar encounters \citep{larwood,quillen}.
Alternatively, the Sun could have been part of a wide binary which was
later disrupted (e.g., as \citealt{furlan}).

As shown by \citet{larwood} in their study of the Beta Pictorus system,
and others (e.g. \citealt{kobayashi})
a close stellar encounter can scatter planetesimals.
For an equal mass parabolic encounter with pericenter distance $q$,
planetesimals with semimajor axis $a \gtrsim q/2$ are scattered, 
whereas those with $a \lesssim q/2$ are uniformly perturbed into higher
eccentricity and inclination orbits \citep{larwood,kobayashi}.
This division depends on the timescale of the encounter
compared to the orbital rotation period of the planetesimals.
In the inner disk, the planetesimals rotate quickly compared to the encounter
time and the collision can be treated with a secular approximation, averaging
over the entire orbit,  as done in the appendix by \citet{kobayashi}.
However, objects with larger semimajor axes are moving at 
speeds similar to the encountering star (assuming a parabolic orbit). 
Consequently the resulting perturbation on outer planetesimals
is strongly dependent on their longitudes, as shown explicitly 
by \citet{larwood}.
If the collision is very fast compared to the rotational timescale 
of the planetesimals, 
as would be the case if the star was on a highly hyperbolic orbit,
then the impulse approximation can be used to estimate 
the perturbations caused by the encounter (e.g., \citealt{eggers,pfalzner}).

Previous work considering the effect of a stellar encounter on
our solar system has considered parabolic orbits
with pericenter of size 100-200 AU \citep{ida,kobayashi}.
At this distance the velocity of
the perturber at pericenter is similar to the rotational velocity
of KBOs.  For such an encounter the Kuiper belt is in a transition region. 
The KBOs experience increases in mean eccentricity and inclination 
which are dependent on their initial semimajor axes. 
Because the final KBO orbits in this region 
also depend on the initial longitude,
the eccentricity and inclination distribution broadens following
the encounter.
\citet{ida, kobayashi} restricted their exploration to fairly
large stellar pericenters $q>100$ AU, which minimized
the eccentricity variations in the KBOs caused by the encounter.
\citet{ida,kobayashi} suggested that Neptune's migration
followed the encounter.
KBOs with initial eccentricities above 0.15 are unlikely to
be captured in the 3:2 mean motion resonance with a migrating Neptune 
\citep{malhotra95,hahn1999}.  
The encounters considered by \citet{ida,kobayashi} 
were consistent with a scenario that included migration
of Neptune since the encounters
did not induce high KBO eccentricities.

In this paper we extend the previous explorations by \citet{ida,kobayashi} 
to include stars on orbits with smaller pericenters.  
In this regime the planetesimal perturbations 
are strongly dependent on the planetesimal longitude during the 
encounter.  
Consequently we expect a larger spread in the resulting planetesimal 
planetesimal orbital properties, as seen in the outer parts of the planetesimal
disks simulated by \citet{larwood,ida,kobayashi}.
Here we explore a regime where the Kuiper Belt is more effectively
scattered by the encounter than previously works have considered.


%

\section{Constraints on parameter space of the encounter}

In this section we discuss preliminary constraints on the orbit, mass
and initial velocity of the stellar impactor.  We maintain our hypothesis that
the Kuiper Belt is strongly scattered by the encounter, but that it must retain
low eccentricity and inclination objects as are found
in the classical Kuiper Belt.  Furthermore we assert 
that Uranus's orbit is not strongly perturbed by the encounter.

\begin{figure*}
\epsscale{1.50}
\plotone{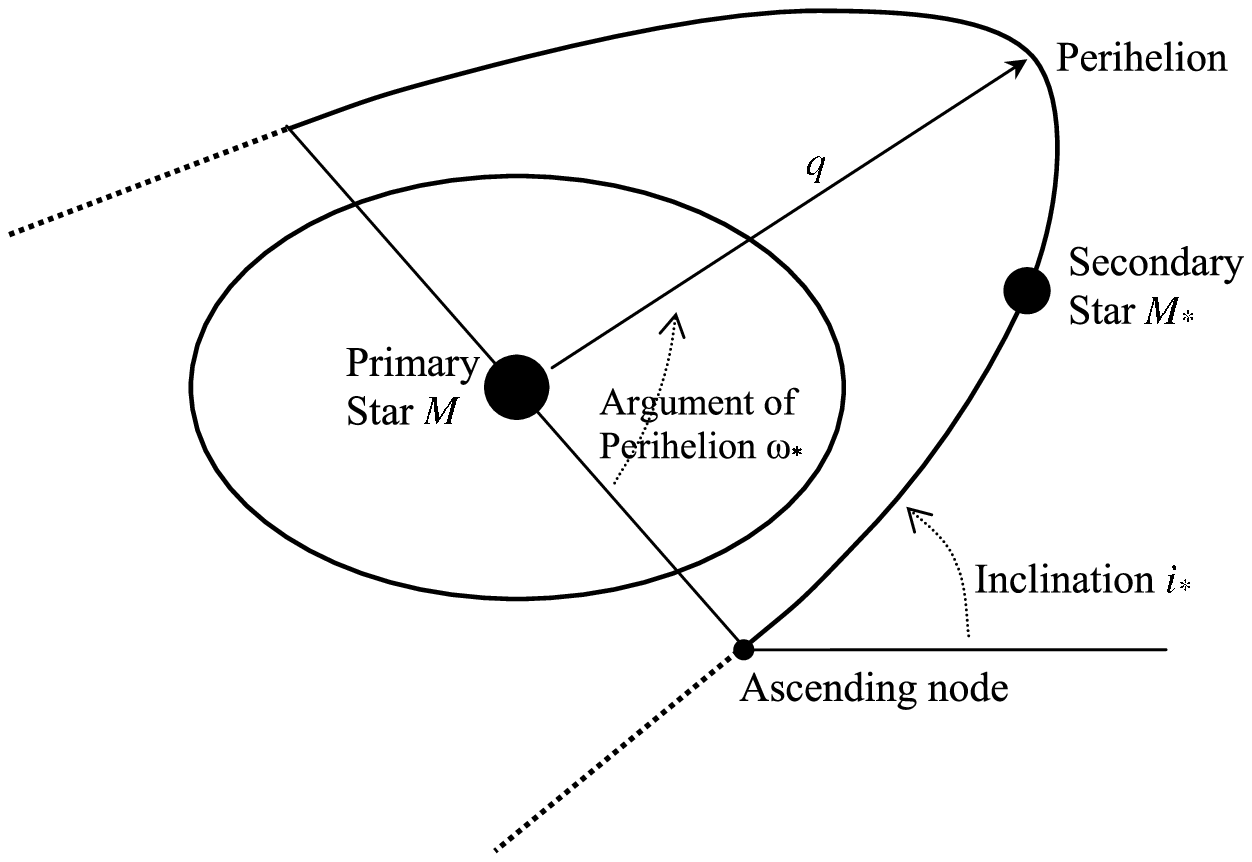}
\figcaption{
\label{fig:impact}
Orbital angles for parabolic and hyperbolic encounters, based
on the similar figure by \citet{kobayashi}.
\vskip 5truein
}
\end{figure*}

The Sun/stellar encounter is described by 5 free parameters,
the mass of the star compared to the Sun, 
$M_*/M_\odot$, the velocity of the star 
distant from the Sun, $v_\infty$, the distance of closest
approach from the Sun or perihelion of the star, $q$, the inclination 
of the star's orbit with respect to the ecliptic, $i_*$, and the argument
of perihelion, $\omega_*$. For a pictorial view
of these angles see Fig. \ref{fig:impact}  which is
based on Fig.~1 by \citet{kobayashi}.
The velocity of the incoming star 
is related to the eccentricity of the orbit $e_*$, where
$v_\infty^2  = (e_*-1) \left( { G(M_\odot + M_*) \over q}\right)$ and 
$e_*>1$ for a hyperbolic orbit.

For a star to scatter a planetesimal effectively, the star's velocity 
must be greater than the circular velocity of the planetesimal.
The velocity of a star in a hyperbolic orbit at perihelion is 
\begin{equation}
v_q^2 =  (1 + e_*){ G(M_\odot + M_*) \over q}.
\end{equation}
We compare the star's velocity at perihelion with the velocity, $v_c$, 
of a particle in a circular orbit with semi-major axis $a$;  
$v_c = \sqrt{GM_\odot/a}$.
Allowing Neptune to be in the scattered region, we require 
$v_c \gtrsim v_q$ at Neptune's $a=30$ AU which implies that
\begin{equation}
q/\sqrt{1+e_*} \lesssim 30 {\rm AU}.
\end{equation}
This sets pericenter $q \lesssim 50$AU if the star 
is on a parabolic orbit and if Neptune is in the scattered region.

We now constrain $e_*$ based on the most probably stellar incoming velocity.
Collisions are unlikely from field stars, so we expect an incoming
stellar velocity of order a few km/s, typical of the velocity
dispersion in young star clusters which have not yet dispersed or disrupted.
\begin{equation}
v_\infty^2  = (e_*-1) \left( { G(M_\odot + M_*) \over q}\right)
\end{equation}
The circular velocity is 4.2 km/s at a semi-major axis of 50~AU, of
order that which might be found in a young unbound cluster.
Since we don't expect $v_\infty$ to be greater than a few km/s if
the perturber is in a birth cluster,  we infer that
the encounter was likely to be nearly parabolic, ($e_*\lesssim 2$).  
A hyperbolic orbit with a small pericenter
from a faster field star would be possible, but it is
much less likely than a parabolic one from a star 
in a dense birth cluster.
If the encounter was a result of an interaction with a binary,
then the velocity of the encounter would
depend on the nature of the binary and how it was disrupted.

We can set a condition on the orbit of the impactor by specifying that
Uranus does not significantly increase in inclination or eccentricity following
the encounter.  Uranus is at small enough semimajor axis that we may assume
that it moves quickly compared to the timescale of the encounter.
To estimate the affect of the encounter on Uranus
we use equations 15 and 16 from \citet{kobayashi} which are appropriate
for parabolic encounters in a secular approximation.
\begin{equation}
i \approx {3 \pi \over 8 \sqrt{2} }
{M_*/M_\odot \over \sqrt{ 1 + M_*/M_\odot}}
\left({ a \over q}\right)^{3/2} \left| \sin 2  i_* \right|
\end{equation}
Based on Uranus's inclination, $0.77^\circ$, and semimajor axis, $19.2$ AU,
and assuming an encounter with $q \sim 50$AU we find 
\begin{equation}
M_* \sin 2 i_* \lesssim 0.087.
\end{equation}
If the star was low mass $M_* \sim 0.1$, then a reasonably
large range of inclinations is allowed.
Since Uranus's eccentricity increase
is $\propto (a/q)^{5/2}$ \citep{kobayashi},
Uranus's low eccentricity is less of a constraint than
its low inclination.
If Uranus was not strongly perturbed by the encounter, 
then the stellar 
impactor was likely to be a low mass star or on a low inclination orbit.


For particles with semi-major axes similar to the pericenter distance
of the encounter, the perturbation caused by the encounter
is strongly dependent upon the particle's longitude.
Since these particles move slower than the star, we can use
the impulse approximation to estimate the size of the 
perturbations.
Each particle undergoes a change in its velocity of order
\begin{equation}
\left| \Delta { V} \right| 
  \sim  
  { 2 G M_* \over b v_q}
\end{equation}
where $b$ is the distance between the object and the star's closest approach to it,
and $v_q$ is the velocity of the star at pericenter 
(e.g., \citealt{B+T,eggers}).
The velocity change is in the direction toward the star at its position of
closest approach.
Using cylindrical coordinates ($R,z$) where $R$ is in the ecliptic,
the position of the star's pericenter is 
$R = q \sqrt{ \cos^2 \omega_* + \sin^2 \omega_* \cos^2 i_*}$, 
$z = q \sin \omega_* \sin i_*$.
For an object opposite pericenter   ($180^\circ$ away)
\begin{equation}
{\Delta v_z \over v}  \sim \sin i \sim 
\left({M_* \over M_\odot} \right)^{1/2}
{\sqrt{2 a q} \over a+q}
\sin \omega_* \sin i_*
\end{equation}
To allow the inclination of particles opposite pericenter
to remain low, we restrict the encounter
to low arguments of perihelion and low inclinations.

\section{Simulations of hyperbolic encounters}

From the previous section we found that if Neptune is in 
the scattered region, then $q\lesssim 50$AU. If the incoming star
is part of the Sun's birth cluster then we expect $e_* \lesssim 2$.  
To maintain
Uranus's low eccentricity and inclination, we require low inclination
orbits and low mass stars.  To ensure that the Kuiper Belt retains a low
eccentricity and inclination population we require orbits with
low inclinations and low arguments of perihelion.
We take these estimates as guidelines for our numerical integrations.

To investigate the actual affect of a such a close stellar encounter 
on a primordial cold Kuiper belt,
we numerically integrated the two body Sun/stellar gravitational system.
Our code is a conventional Burlisch-Stoer numerical scheme which
considers only the force from Newtonian gravity.
Simultaneously during the encounter 
we integrate 1000 low mass particles (planetesimals) per run 
which were initially
placed in circular orbits in the ecliptic.
These particles are influenced by the gravitational force from the Sun
and star but do not themselves act on the Sun or star.
The integrations were begun with the star at 500 AU from the Sun,
and ended when the star reached an equivalent distance from the Sun
following the encounter.
We began our simulations with planetesimals 
distributed randomly in semi-major axis between 30 and 70AU and
randomly chosen mean longitudes.
Following the encounter, the final semi-major axes, eccentricities
and inclinations of the particles were tabulated.

Based on our constraints from the previous section,
we explored planetesimal eccentricity and inclination distributions for
encounters spanning $0.1< M_*/M_\odot < 0.5$, $40 < q < 70$ AU with differing
orbital inclinations and arguments of perihelion 
($i_*, \omega_*$).  Since we expect
that the encounter could have been in the Sun's birth cluster,
we restrict the incoming stellar velocity $v_\infty$ to less than
a few km/s, equivalent to nearly parabolic encounters, $e_* <2$.
In this region of parameter space, we searched for final planetesimal
distributions which contained both scattered and classical 
type KBO population analogs.

\begin{figure*}
\epsscale{1.50}
\plotone{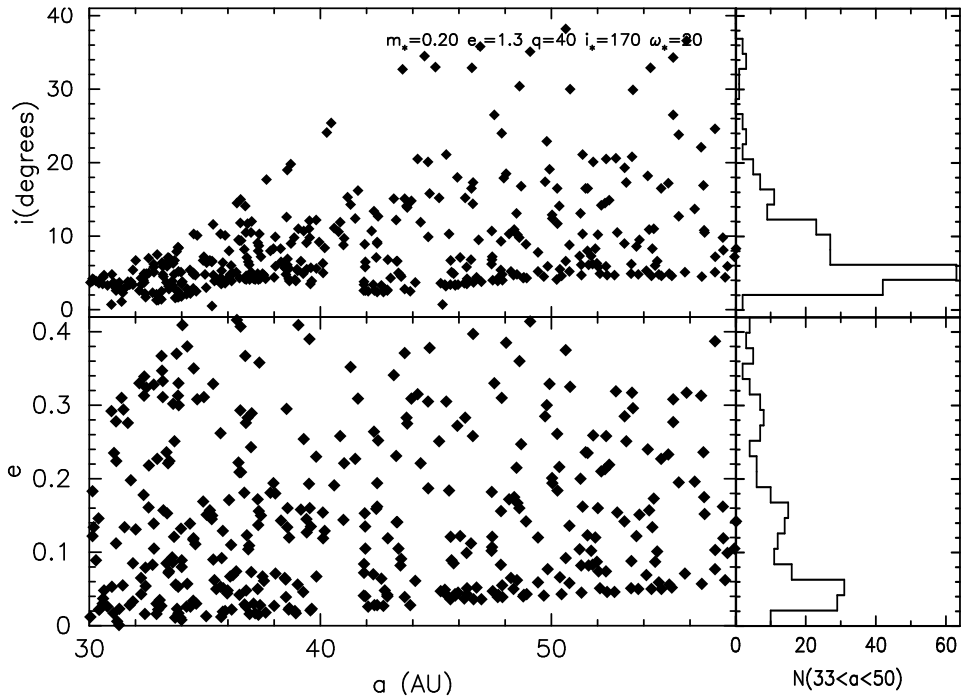}
\figcaption{
\label{fig:retrograde}
Planetesimal distribution following a retrograde hyperbolic encounter with
a star.  The star has $M_* = 0.2 M_\odot$, orbit pericenter at $q=40$AU,  
inclination $i_* = 170^\circ$, argument of perhelion
$\omega_* = 20^\circ$ and eccentricity $e_* = 1.4$.
The lower left shows the eccentricity distribution as a function of semi-major
axis.  The upper left shows the inclination distribution as a function
of semi-major axis.  The right hand side shows histograms for
planetesimals with semi-major axes between 35 and 50~AU.
This low inclination encounter scatters a significant fraction
of the planestesimals, while retaining a low inclination and low eccentricity
analog to the classical Kuiper Belt population.
}
\end{figure*}

In Fig. \ref{fig:retrograde} we show the planetesimal distribution
following a retrograde encounter with a star of $M_*/M_\odot = 0.2$, 
$e_* = 1.4$, pericenter $q = 40$AU, 
inclination $i_* = 170^\circ$ and argument of perihelion 
$\omega_* = 20^\circ$. 
In Fig. \ref{fig:retrograde}, histograms are also shown for all particles
with final semi-major axes between 33 and 50~AU.
Examination of this figure 
shows that the inclination and eccentricity
distribution following the encounter is wider at larger semi-major axes
as expected from a comparison between the timescale of the encounter with
the orbital period of the objects.
We see that a significant fraction, but not all, of the
belt is scattered, 
indicating that the final eccentricity and inclination of an 
object depends on its longitude during the collision (as shown
explicitly by \citealt{larwood}).
About 30\% of objects with final semimajor axes between 33 and 50 AU
have inclinations above $10^\circ$ and eccentricities above $0.1$. 
However, low eccentricity ($e < 0.05$)
and low inclination ($i<6^\circ$) objects remain in the 
planetesimal distribution.  

\begin{figure*}
\epsscale{1.50}
\plotone{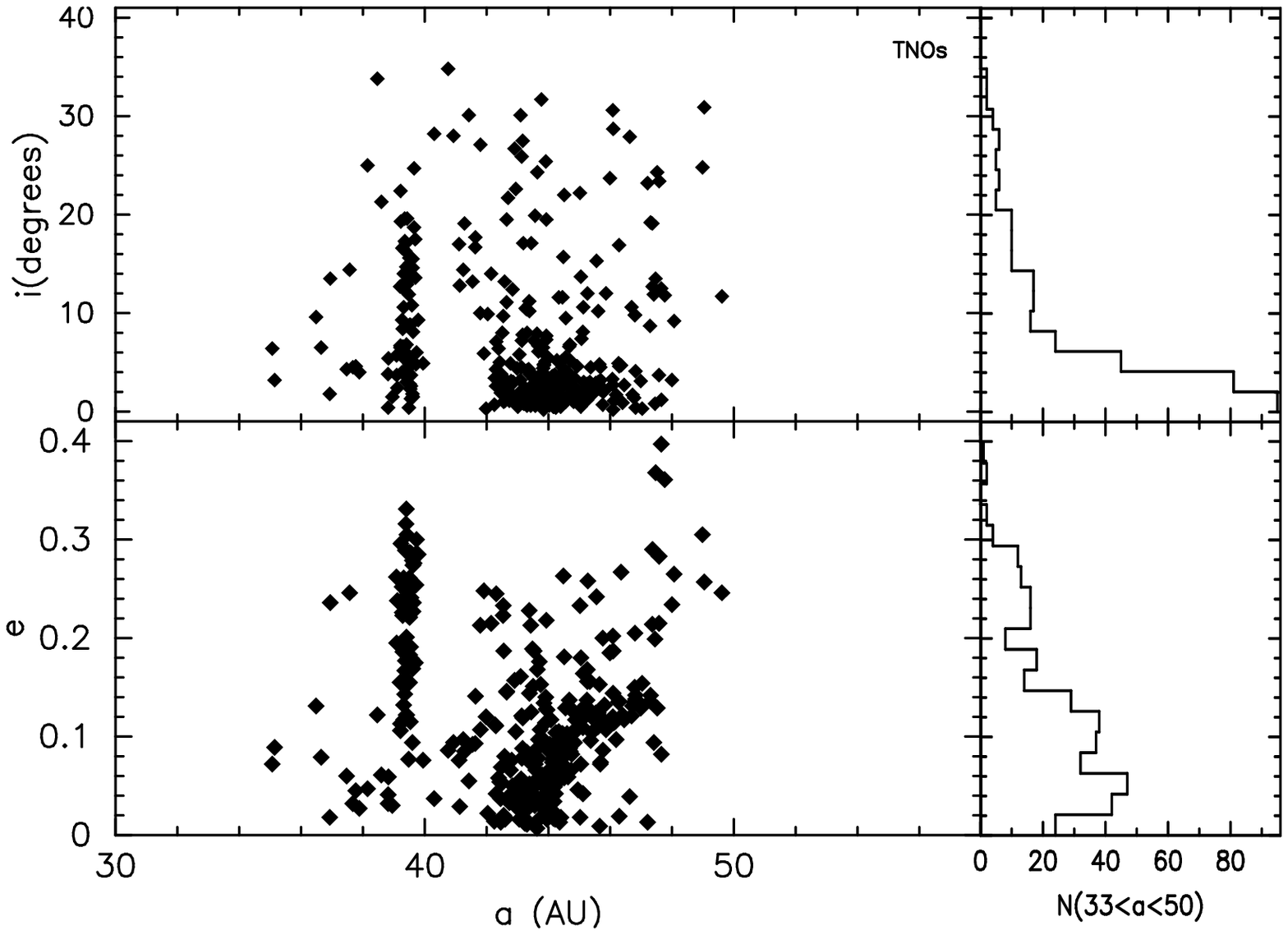}
\figcaption{
\label{fig:TNOs}
Known Trans-Neptunian objects from the minor planet center database
as of Jan 7, 2004. 
To ensure accurate orbits, only objects that have been observed
over more than one opposition have been plotted.
The known KBO distribution 
is displayed identically to the simulation shown in Figure \ref{fig:retrograde}.
The known KBO distribution of orbital elements 
has high eccentricity and inclination tails similar to those seen
in the simulation presented in Fig.~\ref{fig:retrograde}.
The known KBO distribution is affected by observational
selection affects and has had 4 billion years of dynamical evolution
since formation.  High eccentricity objects that are not in mean
motion resonances
with Neptune are likely to have been removed from the Kuiper belt.
}
\end{figure*}

Fig. \ref{fig:retrograde} can be compared to the distribution
of known Trans-Neptunian objects 
(as listed by the Minor Planet Center
Jan. 7, 2004 \footnote{http://cfa-www.harvard.edu/iau/lists/TNOs.html})
which is given in the same format in Fig. \ref{fig:TNOs}.
To ensure accurate orbits, only objects that have been observed 
over more than one opposition have been plotted. 
We must keep in mind that the known population of KBOs is likely to be biased
by the systematics of the observations (e.g., as discussed
by \citealt{bernstein,brown}).  The KBO has also had $\sim$ 4 billion
years to evolve dynamically following the hypothetical collision considered
here (e.g., as explored by \citealt{kuchner}).

While the simulation of the retrograde encounter 
shown in Fig. \ref{fig:retrograde} 
does not exactly match the known KBO population 
(the most obvious mismatch being the radial distribution), 
it successfully predicts the combined presence of 
objects with high inclinations and eccentricities while retaining  
a population with low inclinations and eccentricities,
as required by the known KBO distribution.  
The inclination and eccentricity distribution
following the encounter have high inclination and eccentricity tails,
similar to those seen in the KBO comparison histograms.
An object at the location of Neptune (30AU) 
can remain in a low eccentricity and inclination orbit, so Neptune
itself could have survived the encounter with minimal changes in
its orbital elements.  

\begin{figure*}
\epsscale{1.50}
\plotone{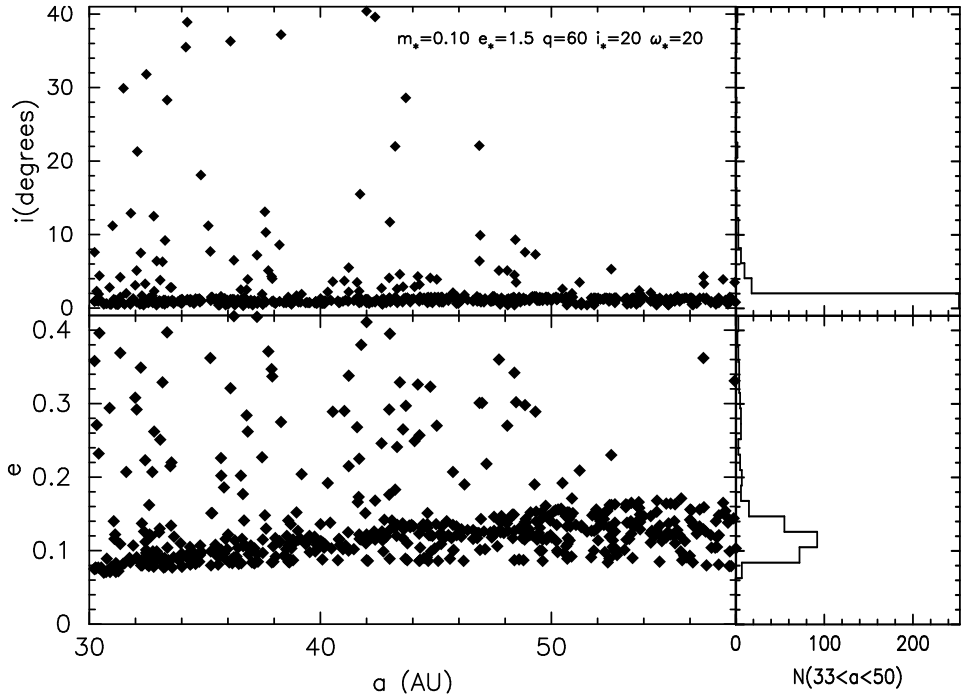}
\figcaption{
\label{fig:prograde}
Similar to Figure \ref{fig:retrograde} except the encounter is prograde. 
This simulation corresponds to an encounter with $M_* = 0.1M_\odot$,
$e_* = 1.5$, $i_* = 20^\circ$, $\omega_* = 20^\circ$, and $q_0 = 60$AU.
Prograde encounters can also cause a dichotomy in the scattered
population similar to what is seen in the Kuiper Belt.
However, we have failed to identify a region in parameter
space that retains low eccentricity objects as well as low
inclination ones, and still scatters significant fraction of the belt.
}
\end{figure*}

The distribution of orbital elements for the retrograde encounter 
can be contrasted, for example,
with the prograde encounter shown in 
Fig. \ref{fig:prograde} which does not exhibit 
as broad a distribution of eccentricities and inclinations.
The simulation (Fig. \ref{fig:retrograde}) we chose to directly compare to the current
KBO population (Fig. \ref{fig:TNOs}) corresponded to 
a low inclination retrograde encounter.
A prograde encounter can more effectively perturb the disk because
the collision effectively lasts longer in the frame
rotating with each planetesimal.
We also find prograde collisions which are capable of scattering
the disk, and an example of one is shown in Fig. \ref{fig:prograde}. 
While low inclination retrograde encounters fail to produce
very low inclination objects in the belt,
we find that the low inclination prograde close encounters fail to produce 
low eccentricity objects,
requiring subsequent evolution to account for them, including 
the low eccentricity of Neptune itself.
In addition, we have had difficulty identifying a region
of parameter space in which a low
eccentricity and inclination population is retained but a
a significant fraction of the belt is scattered.

We consider what effects subsequent evolution could have on the resulting
planetesimal distribution shown in Fig. \ref{fig:retrograde} for the retrograde
encounter.
Scattered objects at high eccentricities that were not placed in
resonances with Neptune would be quickly removed from the system
because they cross Neptune's orbit and so could be scattered by Neptune.
High eccentricity objects at the location of Pluto (40AU) would be removed,
leaving a population of Plutinos which include
high eccentricity and inclination objects.  Alternatively
because there remains a population of low eccentricity objects, 
subsequent migration by Neptune can resonantly capture objects
into the 3:2 resonance accounting for the Plutinos.   Because
the low eccentricity objects include objects at high inclinations,
the resulting captured
Plutinos would also include high inclination objects.

The simulation shown in Fig. \ref{fig:retrograde} 
does not necessarily require Neptune
to migrate to account for the Plutinos since non resonant high
eccentricity objects at 40 AU will be removed by subsequent 
scattering by Neptune.  
If we consider models which include Neptune's migration,
then a closer (smaller pericenter) encounter 
than simulated in Fig. \ref{fig:retrograde}  would be required.
This follows because low eccentricity objects captured at smaller
semi-major axes must include high inclination objects such
as are found past 40 AU in this simulation.

While no objects are found at extremely 
low inclinations in the retrograde simulation,
(Fig. \ref{fig:retrograde}),
subsequent evolution may broaden the inclination distribution, 
reducing the inclinations of some objects.
We began the simulations with zero inclination and 
eccentricity objects.  Had we started the simulation
with a distribution of inclinations and eccentricities, the final
distribution would have included lower inclination objects.
If the stellar orbit has a higher inclination than the simulation
shown in \ref{fig:retrograde}, then following the encounter
the mean inclination is higher, however
more objects remain at low eccentricities, and the eccentricity
distribution is narrower.

\begin{figure*}
\epsscale{1.50}
\plotone{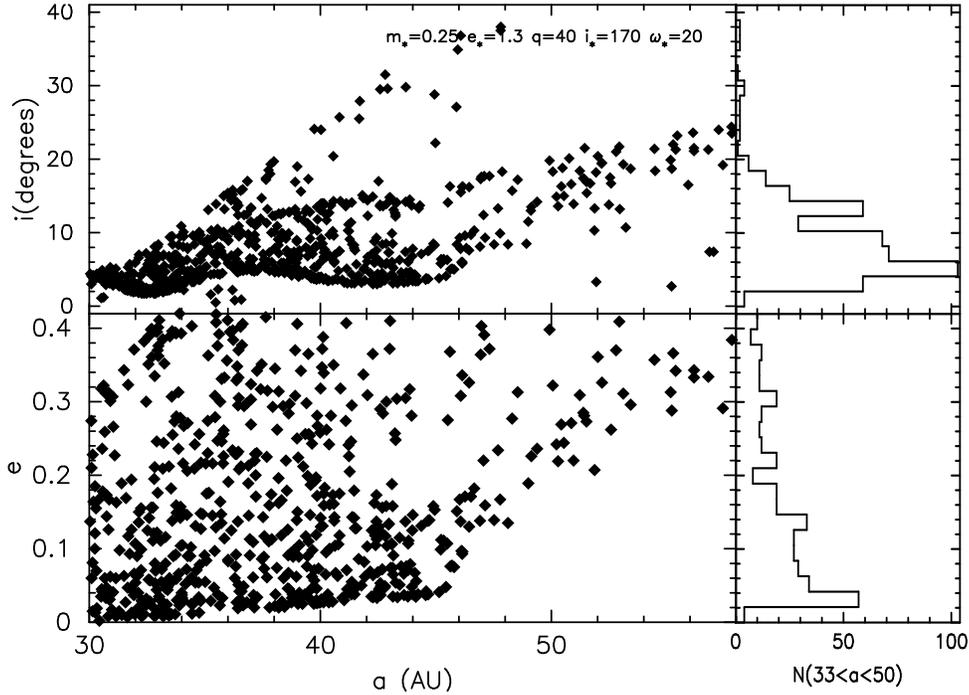}
\figcaption{
\label{fig:truncated}
Similar to Figure \ref{fig:retrograde} except the planetesimal disk prior
to the encounter was truncated at 45 AU prior to the encounter. 
This simulation corresponds to an encounter with $M_* = 0.25M_\odot$,
$e_* = 1.3$, $i_* = 170^\circ$, $\omega_* = 20^\circ$, and $q_0 = 40$AU.
Following the encounter, objects initially near the
edge of the disk can be scattered to higher eccentricity, and semi-major
axis orbits.   Past 45 AU, both the mean eccentricity and inclination
increase with semi-major axis.  A previously truncated disk cannot
explain the eccentricity increase with semimajor axis seen in the Kuiper
belt because there is no increase in mean inclination past 45 AU in
the Kuiper belt.
}
\end{figure*}

Deep observational searches 
suggest that the Kuiper Belt is truncated at 50 AU \citep{allen}
and does not contain a large population of "cold" or near zero eccentricity
objects \citep{allen2002}.
Our simulated encounters do not truncate the disk, but instead
increase the radial distribution of any previously defined edge
(e.g. as shown by \citep{larwood} in similar simulations for the Beta Pic
disk).  It is possible that an edge in the KBO distribution existed
prior to the stellar encounter.  To investigate the effect of this edge,
we ran encounter simulations with a  
a predefined edge at 45 AU in the planetesimal
distribution prior to the encounter.
Following the encounter, the planetesimal distribution 
is shown in Fig. \ref{fig:truncated}. The simulated 
encounter was retrograde with $M_* = 0.25M_\odot$,
$e_* = 1.3$, $i_* = 170^\circ$, $\omega_* = 20^\circ$, and $q_0 = 40$AU.
The resulting planetesimal distribution 
exhibits a rise in both mean eccentricity and inclinations with
semi-major axis for objects past 45 AU.  A rise in mean eccentricity
is exhibited by the known KBOs shown in Fig. \ref{fig:TNOs}, however
the known KBO distribution lacks an increase in mean inclination 
in the region same region.   
Consequently the observed increase in mean eccentricity with semimajor
axis seen in the KBO population is unlikely to be 
explained by an encounter perturbing
a pre-existing disk edge.  
Since the brightness of an object
depends on its distance from the Sun to the fourth power, high
eccentricity KBO's are 
more likely to be discovered when they are near perihelion,
even taking into account the time they spend at small distances
from the Sun compared to that spent at large distances.
This selection effect may have reduced the number of known
low eccentricity KBOs objects at large semimajor axis.

\section{Discussion and Summary}

In this paper we have numerically investigated the effect of
a close stellar encounter on the primitive Kuiper Belt.  We have
extended our study to collisions with lower mass stars and
smaller pericenters than have the pioneering studies of 
\citet{ida,kobayashi}.
We find that low mass, low inclination encounters can produce
a dichotomy in the resulting planetesimal distribution, containing
both scattered objects with high eccentricities or high
inclinations as well as low inclination and low eccentricity
objects, similar to what is seen in the Kuiper Belt.  In particular
a $ 0.2 M_\odot$ star passing with a pericenter of 45 AU from the
Sun, in a retrograde, highly inclined orbit produces inclination
and eccentricity distributions with high end tails,
similar to those seen in the KBO distributions.

To more thoroughly test such a scenario it is necessary to numerically
evolve the scattered population for the lifetime of the solar system
as done for the classical KBO population by \citet{kuchner}.
High eccentricity objects at 40AU which were not placed
in resonance with Neptune are likely to be removed from
the Kuiper Belt by scattering with Neptune.  
Consequently the 3:2 resonance
represents a region of stability; objects placed
in this resonance could stay there for the lifetime of the solar system.
Objects placed in this resonance by the encounter
could explain the high inclination Plutino population.  
Objects currently residing in resonances that have low capture
probabilities such as the 5:2 resonance \citep{chiang} might
be more naturally explained with a scattering scenario rather than
a Neptune migration scenario.  Alternately one could regard the
existence of such objects as a consequence or
prediction of the impact scenario explored here.
Since long lived regions of dynamical stability
differ from those selected by resonance capture, study of future 
measurements of the
of KBO orbital distribution may support or reject a scenario
which includes a close stellar encounter.

Low eccentricity objects remain in the belt following the encounter,
which implies that there remains a population which 
can be captured into resonance by subsequent migration by Neptune 
(e.g., \citealt{nagasawa2000}),
however Neptune's migration may not be needed to explain the highest
Plutino eccentricities.  This may reduce the extent of
migration required by recent migration models \citep{gomes,levison}.
\citet{hahn2003} have suggested that the resonant KBO
populations might be better explained
by Neptune's migration into a previously ``hot'' Kuiper Belt, a situation
which could occur after the encounter we have simulated here.

Following the stellar encounter, we expect many objects to be removed
from the belt by dynamical evolution, suggesting that there was 
a period of time when short period comets were produced in
great numbers.  One interesting possibility
is that this epoch corresponded to the period
of ``Late Heavy Bombardment'' when the solar system 
was a few $\times 10^8$ years old. This possibility would require 
a cometary origin for the lunar impacts 
associated with the Late Heavy Bombardment Era  (for further
discussion on the cometary vs. asteroidal origin see 
\citealt{levison2000,levison2001}). 
Because of the large mass of material likely to be sent into
the inner solar system by Neptune, this epoch
is also very likely to be the time during which Neptune migrated.
Because of the narrowness of Neptune's mean motion resonances,
a close stellar encounter probably did not occur after the
migration of Neptune.   This follows because such an encounter
would have effectively removed objects from resonance.
The encounter explored here would also place objects in the 
2:1 resonance with Neptune, and so may not naturally account
for the low fraction of Twotinos compared to Plutinos unless there
was a pre-existing edge to the Kuiper Belt.

While a close encounter with a low mass star could account
for the inclination distribution and dichotomy of orbital
properties seen in the Kuiper Belt, as seen
in our simulations, it does not truncate the disk.
Because they pull tidal arms outward, stellar encounters do not
tend to sharply truncate disks 
(e.g. \citealt{pfalzner}).  If the Kuiper Belt contains an edge at 50~AU
as suggested by \citep{allen}, then it is much more likely
to have been caused by a companion, either stellar or planetary.
We note that the encounters simulated here did remove $\sim 10\%$ 
of the planetesimals in the region 30--70AU objects from the Solar system.
The encounter could have ejected a primordial planet at a semimajor
axis of $\sim 50$ AU, leaving only its signature behind in the KBO distribution.

If a stellar encounter did perturb the young Kuiper Belt, then
the limits on the likely stellar birth cluster of the Sun can
be somewhat relaxed from the lower limits placed by \citet{adams},
as suggested by \citep{fuente_marcos}.
In particular a higher stellar density birth cluster, similar to that
in the Orion nebula with a few times 
$10{^4}$ stars pc$^{-3}$ \citep{hillenbrand},
or a longer lasting open cluster,
would be required to make the hypothetical collision explored
here probable during the lifetime of the cluster.
The probability $P$, of an object of cross section $\sigma$ undergoing 
a collision while moving through a stellar population of number 
density $n$, with velocity dispersion $v \sim 1$ km/s during time $t$ is
$P = n v \sigma t$ (e.g., \citealt{adams}).
Due to gravitational focusing,
the impact parameter $b$, of an encounter is larger than
the pericenter distance $q$.  For an encounter with
a small pericenter (when $\sqrt{G(M_\odot + M_*)/q} > v_\infty$), 
the impact parameter is 
${b \over q} \sim {\sqrt{G (M_\odot + M_*)/q} \over v_\infty}$.
For an encounter with the Sun 
with pericenter $q\sim 50$~AU,  and $v_\infty = 1$km/s,
we estimate a cross section 
$\sigma = \pi b^2 \sim 3 \times 10^4 AU^2$.
In the Sun's birth cluster the probability that such
an encounter occurred during time $t$
\begin{equation}
P \sim 0.16
\left({ n\over 2 \times 10^4 {\rm pc}^{-3}}\right)
\left({ v\over 1 {\rm km~s}^{-1}}\right)
\left({ \sigma \over 3\times 10^4 {\rm AU}^{2}}\right)
\left({ t \over 1{\rm 10 Myr}}\right). 
\end{equation}
We therefore find that a collision proposed here 
would have been only a moderately improbable event (15 percent level)
if the Sun were born in a dense cluster similar
to that in the Orion nebula.  Here we have only considered
the probability of a close collision and have not
restricted the orientation of the encounter
with respect to the ecliptic.  Had we taken into account
the narrow region of allowed orbital inclinations and arguments
of perihelion, our proposed encounter would be about 100 times 
less probable.  The formation time of the KBOs ($\gtrsim 10^8$ years) 
is longer than that used in the above estimate, 
\citep{kenyon98,stern97}, suggesting  
that the Sun resided in a longer lived 
and more massive birth cluster, such as an open cluster.  A longer
timescale would allow the close encounter to occur after the formation
of the KBOs.
Alternatively the Sun could have been part of a wide binary, and
the close encounter could have occurred when the binary was disrupted
(e.g., one of the components
of T Tauri has just recently been disrupted, see \citealt{furlan}).

Currently there are number of examples of young systems which show
clear evidence of external perturbers, including the AeBe stars
HD 141569A which has spiral
structure and asymmetries excited by its binary companion HD 141569B,C 
\citep{augereau2003,quillen2004}, and
HD 100546 which exhibits spiral structure 
which could have been excited by star at a pericenter of about 500 AU 
\citep{quillen2004,grady}.
The tilt in Bet Pictorus's disk has been explained with 
an stellar encounter with a similar sized pericenter \citep{larwood}.
Structure seen in submillimeter emission of older systems have recently
been interpreted in terms of high eccentricity outer planets. 
For example models for the submillimeter distribution for Vega include an 
outer high eccentricity Jupiter
mass planet \citep{wilner} and that for Epsilon Eridani include 
a Neptune mass moderate eccentricity planet \citep{quillen}.
While the hypothetical encounter with our solar system 
explored here did not significantly increase
Neptune's eccentricity, it easily could have, had the incoming
star approached the Sun from a different
orientation. This suggests 
a possible scenario which could account
for the observed morphology of the the dusty debris disks which
require outer eccentric planets.   
Future study may determine where and when close stellar encounters are likely
to affect young solar systems as well as test and further explore
the possibility that such an encounter occurred in
our own Solar system. 

\acknowledgments
We thank S. Thorndike, E. Chiang, E. Ford, 
P. Varniere, and A. Frank for helpful discussions.
This material is based on upon work supported by the National Aeronautics and 
Space Administration under Grant No.xxx issued through the Origins of
Solar Systems program.  This research was supported in part 
by the National Science Foundation to the KITP 
under Grant No. PHY99-07949.

\clearpage

\clearpage

\end{document}